\documentclass{webofc}
\usepackage[varg]{txfonts}   
\usepackage{siunitx}
\usepackage{natbib}

\pdfoutput=1

\begin{document}
\title{NIKA2 observations of dust grain evolution from star-forming filament to T-Tauri disk:  Preliminary results from NIKA2 observations of the Taurus B211/B213 filament
}
%
%

\author{%
  \lastname{Q.~Nguyen-Luong}\inst{\ref{AUP},\ref{CEA}}
  \and  R.~Adam \inst{\ref{OCA}}
  \and  P.~Ade \inst{\ref{Cardiff}}
  \and  H.~Ajeddig \inst{\ref{CEA}}
  \and  P.~Andr\'e \inst{\ref{CEA}}
  \and  E.~Artis \inst{\ref{LPSC},\ref{Garching}}
  \and  H.~Aussel \inst{\ref{CEA}}
  \and  A.~Beelen \inst{\ref{LAM}}
  \and  A.~Beno\^it \inst{\ref{Neel}}
  \and  S.~Berta \inst{\ref{IRAMF}}
  \and  L.~Bing \inst{\ref{LAM}}
  \and  O.~Bourrion \inst{\ref{LPSC}}
  \and  M.~Calvo \inst{\ref{Neel}}
  \and  A.~Catalano \inst{\ref{LPSC}}
  \and  M.~De~Petris \inst{\ref{Roma}}
  \and  F.-X.~D\'esert \inst{\ref{IPAG}}
  \and  S.~Doyle \inst{\ref{Cardiff}}
  \and  E.~F.~C.~Driessen \inst{\ref{IRAMF}}
  \and  G.~Ejlali \inst{\ref{Tehran}}
  \and  A.~Gomez \inst{\ref{CAB}} 
  \and  J.~Goupy \inst{\ref{Neel}}
  \and  C.~Hanser \inst{\ref{LPSC}}
  \and  S.~Katsioli \inst{\ref{Athens_obs}, \ref{Athens_univ}}
  \and  F.~K\'eruzor\'e \inst{\ref{Argonne}}
  \and  C.~Kramer \inst{\ref{IRAMF}}
  \and  B.~Ladjelate \inst{\ref{IRAME}} 
  \and  G.~Lagache \inst{\ref{LAM}}
  \and  S.~Leclercq \inst{\ref{IRAMF}}
  \and  J.-F.~Lestrade \inst{\ref{LERMA}}
  \and  J.~F.~Mac\'ias-P\'erez \inst{\ref{LPSC}}
  \and  S.~C.~Madden \inst{\ref{CEA}}
  \and  A.~Maury \inst{\ref{CEA}}
  \and  P.~Mauskopf \inst{\ref{Cardiff},\ref{Arizona}}
  \and  F.~Mayet \inst{\ref{LPSC}}
  \and  A.~Monfardini \inst{\ref{Neel}}
  \and  A.~Moyer-Anin \inst{\ref{LPSC}}
  \and  M.~Mu\~noz-Echeverr\'ia \inst{\ref{LPSC}}
  \and  L.~Perotto \inst{\ref{LPSC}}
  \and  G.~Pisano \inst{\ref{Roma}}
  \and  N.~Ponthieu \inst{\ref{IPAG}}
  \and  V.~Rev\'eret \inst{\ref{CEA}}
  \and  A.~J.~Rigby \inst{\ref{Leeds}}
  \and  A.~Ritacco \inst{\ref{INAF}, \ref{ENS}}
  \and  C.~Romero \inst{\ref{Pennsylvanie}}
  \and  H.~Roussel \inst{\ref{IAP}}
  \and  F.~Ruppin \inst{\ref{IP2I}}
  \and  K.~Schuster \inst{\ref{IRAMF}}
  \and  A.~Sievers \inst{\ref{IRAME}}
  \and  C.~Tucker \inst{\ref{Cardiff}}
  \and  R.~Zylka \inst{\ref{IRAMF}}
  \and  A.~Bacmann \inst{\ref{IPAG}}
  \and  A.~Duong-Tuan \inst{\ref{CEA}}
  \and  N.~Peretto \inst{\ref{Cardiff}}
  \and  A.~Rigby \inst{\ref{Cardiff}}
}
\institute{
     CSMES, 
     The American University of Paris, 2bis, Passage Landrieu, 75007, Paris, France
  \label{AUP}
  \and
  Universit\'e C\^ote d'Azur, Observatoire de la C\^ote d'Azur, CNRS, Laboratoire Lagrange, France 
  \label{OCA}
  \and
  School of Physics and Astronomy, Cardiff University, CF24 3AA, UK
  \label{Cardiff}
  \and
  Universit\'e Paris-Saclay, Université Paris Cité, CEA, CNRS, AIM, 91191, Gif-sur-Yvette, France
  \label{CEA}
  \and
  Universit\'e Grenoble Alpes, CNRS, Grenoble INP, LPSC-IN2P3, 38000 Grenoble, France
  \label{LPSC}
  \and	
  Max Planck Institute for Extraterrestrial Physics, 85748 Garching, Germany
  \label{Garching}
  \and
  Aix Marseille Univ, CNRS, CNES, LAM, Marseille, France
  \label{LAM}
  \and
  Universit\'e Grenoble Alpes, CNRS, Institut N\'eel, France
  \label{Neel}
  \and
  Institut de RadioAstronomie Millim\'etrique (IRAM), Grenoble, France
  \label{IRAMF}
  \and 
  Dipartimento di Fisica, Sapienza Universit\`a di Roma, I-00185 Roma, Italy
  \label{Roma}
  \and
  Univ. Grenoble Alpes, CNRS, IPAG, 38000 Grenoble, France
  \label{IPAG}
  \and
  Institute for Research in Fundamental Sciences (IPM), Larak Garden, 19395-5531 Tehran, Iran
  \label{Tehran}
  \and
  Centro de Astrobiolog\'ia (CSIC-INTA), Torrej\'on de Ardoz, 28850 Madrid, Spain
  \label{CAB}
  \and
  National Observatory of Athens, IAASARS, GR-15236, Athens, Greece
  \label{Athens_obs}
  \and
  Faculty of Physics, University of Athens, GR-15784 Zografos, Athens, Greece
  \label{Athens_univ}
  \and
  High Energy Physics Division, Argonne National Laboratory, Lemont, IL 60439, USA
  \label{Argonne}
  \and  
  Instituto de Radioastronom\'ia Milim\'etrica (IRAM), Granada, Spain
  \label{IRAME}
  \and
  LERMA, Observatoire de Paris, PSL Research Univ., CNRS, Sorbonne Univ., UPMC, 75014 Paris, France  
  \label{LERMA}
  \and
  School of Earth \& Space and Department of Physics, Arizona State University, AZ 85287, USA
  \label{Arizona}
  \and
  School of Physics and Astronomy, University of Leeds, Leeds LS2 9JT, UK
  \label{Leeds}
  \and
  INAF-Osservatorio Astronomico di Cagliari, 09047 Selargius, Italy
  \label{INAF}
  \and 
  LPENS, ENS, PSL Research Univ., CNRS, Sorbonne Univ., Universit\'e de Paris, 75005 Paris, France 
  \label{ENS}
  \and  
  Department of Physics and Astronomy, University of Pennsylvania, PA 19104, USA
  \label{Pennsylvanie}
  \and
  Institut d'Astrophysique de Paris, CNRS (UMR7095), 75014 Paris, France
  \label{IAP}
  \and
  University of Lyon, UCB Lyon 1, CNRS/IN2P3, IP2I, 69622 Villeurbanne, France
  \label{IP2I}
}

\abstract{%
To understand the evolution of dust properties in molecular clouds in the course of the star formation process, we constrain the changes in the dust emissivity index 
from star-forming filaments to prestellar and protostellar cores to T Tauri stars. 
Using the NIKA2 continuum camera on the IRAM 30~m telescope, we observed the Taurus B211/B213 filament at 1.2\,mm and 2\,mm with unprecedented sensitivity and used the resulting maps to derive the dust emissivity index $\beta$. 
Our sample of 105 objects detected in the $\beta$ map of the B211/B213 filament indicates  that, overall, $\beta$ decreases from filament and prestellar cores ($\beta \sim 2\pm0.5$) to protostellar cores ($\beta \sim 1.2 \pm 0.2$) to T-Tauri protoplanetary disk ($\beta < 1$). The averaged dust emissivity index $\beta$ across the B211/B213 filament exhibits a flat ($\beta \sim 2\pm0.3$) profile. 
This may imply that dust grain sizes are rather homogeneous in the filament, start to grow significantly in size only after the onset of the gravitational contraction/collapse of prestellar cores to protostars, reaching big sizes in T Tauri protoplanetary disks.
This evolution from the parent filament to T-Tauri disks happens on a timescale of about 1-2~Myr.
}
\maketitle
\section{Introduction}
\label{sec:intro}
Dense gas in molecular clouds is often (if not primarily) organized along filaments \citep{andre10}. Filaments fragment into dense cores which eventually form stars 
after various evolutionary stages \citep{andre14,marsh16}. Due to the evolution of physical properties from filament to star-forming cores, dust properties such as grain sizes 
and grain composition are expected to change as well. Dust grains can coagulate and grow efficiently in T-Tauri stars and eventually build up planetesimals \cite{ohashi21}. The whole evolutionary process from the parent filament 
to T-Tauri disks is believed to occur on a timescale of $\sim$\,1--2 Myr \citep{andre14}.
Tracing the changes in dust properties is difficult, but models suggest that the dust emissivity index $\beta$ inferred from the ratio of observed intensities $I_1/I_2 $ at two 
(sub)millimeter wavelengths $\lambda_1$, $\lambda_2$ can potentially be used to capture information on the properties of dust grains \citep{Ossenkopf94,Stognienko95,Henning96}. 
If both wavelengths lie in the Rayleigh-Jeans tail of the dust emission spectrum, $\beta \approx \alpha-2$, where $\alpha$ is the observed spectral index 
${\rm log}\, (I_1/I_2) / {\rm log}\, (\lambda_2/\lambda_1)$. 
An accurate estimate of $\beta$ in the (sub)millimeter regime 
is not only crucial for understanding dust properties but also important to derive reliable column densities and masses in molecular clouds. 
For example, the {\it Herschel} Gould-Belt Survey (HGBS) collaboration \citep{andre10} adopted a constant $\beta=2$ to derive column density maps for entire molecular 
clouds (cf. \cite{andre10,palmeirim13,nguyenluong13}). 
However, noticeably lower $\beta$ values have been reported for protostellar cores \citep{galametz19} or T-Tauri stars \citep{Tazzari21}.  

Here, we examine how the dust emissivity index $\beta$ may evolve during star formation  
by focusing on the B211/B213 filament, the most prominent filamentary structure 
actively forming stars in the Taurus molecular cloud (TMC) \citep{goldsmith08,shimajiri19}. 
Owing to its proximity to us (140~pc), cloud substructures (e.g., cores) and young stellar objects (YSOs) at different evolutionary stages have been widely investigated 
in the TMC \citep{hartmann02,hacar13}. 
Compact star-forming dense cores have been detected in the B211/B213 filament using both submillimeter continuum data \citep{marsh16} 
and molecular line (e.g., C$^{18}$O, H$^{13}$CO$^+$) data \citep{onishi02}.
On the larger scale, the filament is believed to be self-gravitating and currently contracting quasi-statically toward its long axis while accreting material from the surrounding environment \citep{palmeirim13,shimajiri19}. Its radial density profile is Plummer-like, with a flat inner plateau $\sim$\,0.1\,pc in diameter, and approaches a $r^{-2}$ power law 
at large radii; meanwhile, the dust temperature profile drops significantly from $\sim\,$15\,K in the ambient cloud to $\sim\,$11\,K 
or less near the filament crest (\cite{palmeirim13}; see also Fig~\ref{fig:betaprofile}). 
Most of these properties were obtained from column density and temperature maps derived from the {\it Herschel} data
assuming a constant $\beta=2$ value.
With NIKA2, we have the 
advantage of being able to take images at 1.2 and 2~mm simultaneously, probing emission in the true Rayleigh-Jeans regime of the dust spectrum.
This allows us to derive a robust $\beta$ index in the entire filament region and to examine variations of $\beta$ between different subregions and at different evolutionary stages.

\begin{figure*}
\begin{tabular}{c}
\includegraphics[width=11.cm]{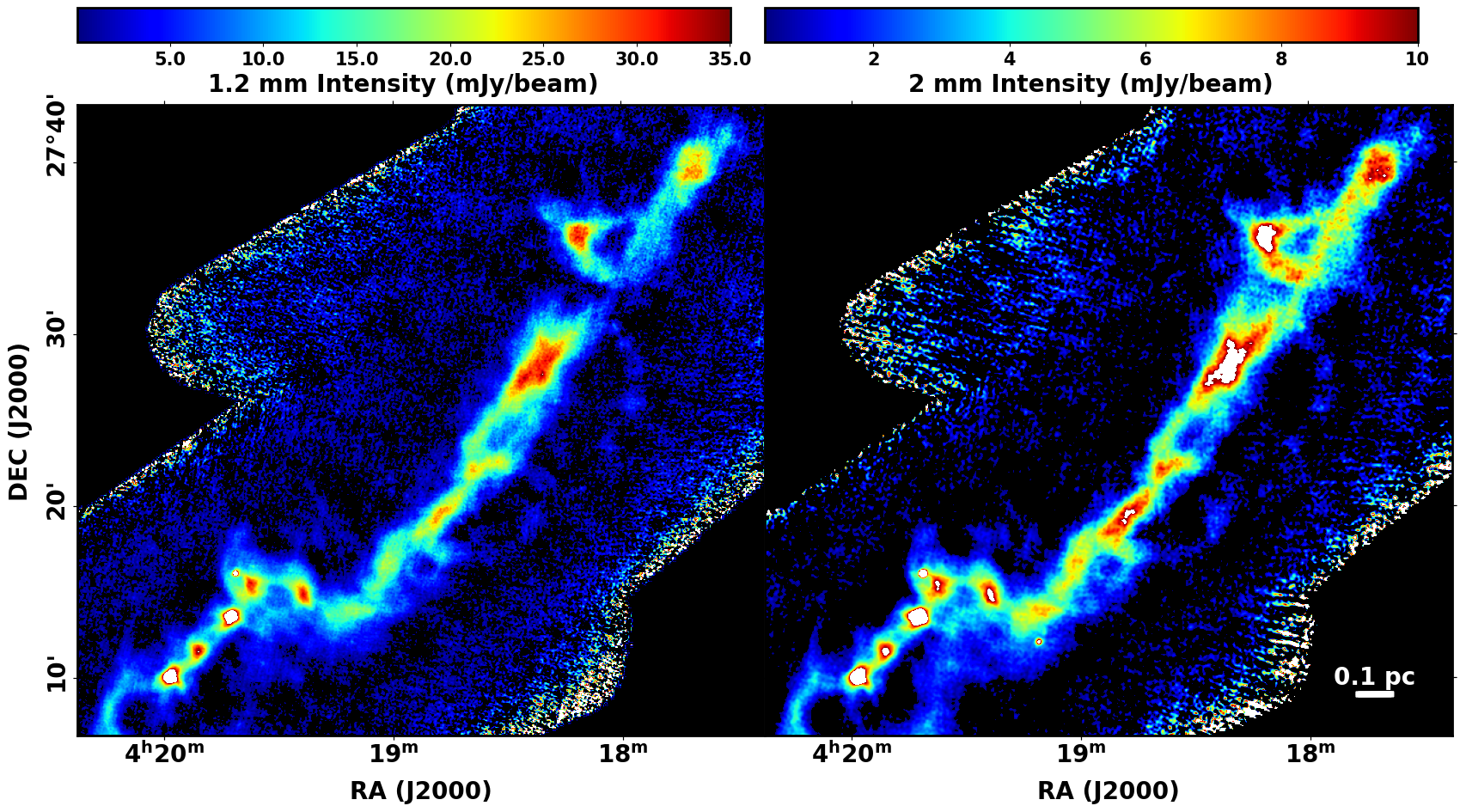}
\end{tabular}
      \caption[]{NIKA2 intensity maps of the Taurus B211/B213 filament region in the 1.2~mm (left) and 2~mm (right) bands.}
\label{fig:Nika_images}
\vspace*{-0.7cm}
\end{figure*}

\section{NIKA2 observations and additional \emph{Herschel} data} 
\label{sec:obs}

The Taurus filament B211/B213 was observed from September 2021 to January 2023 using the New IRAM KID Arrays 2 (NIKA2) camera at the IRAM 30\,m-telescope. This project is part of the 200-hr large program GASTON (Galactic Star Formation with NIKA2; PI: N. Peretto). 
NIKA2 is the new dual-band (150 and 240 GHz) Kinetic Inductance Detectors camera for continuum observations at the 30\,m-telescope \citep[]{Perotto2020}. 
The average beam sizes are 11.1$^{\prime\prime} \pm0.2^{\prime\prime}$ at 1.2~mm and 17.6$^{\prime\prime}\pm0.1^{\prime\prime}$ at 2.0~mm.
In total, we observed 37 hours in good sky conditions with atmospheric opacities ranging from $\tau_{225 \ \mathrm{GHz}}=0.08$ to 0.44 as measured by the IRAM 30-m taumeter. 
The data were reduced using the PIIC/GILDAS pipeline. 
As the resulting two high-sensitivity maps at 1.2~mm and 2.0~mm (Fig.~\ref{fig:Nika_images}) were obtained simultaneously with the same telescope, calibration errors due to differential atmospheric conditions are minimized, and the corresponding ratio map  (Fig.~\ref{fig:betamap}b) is more accurate than otherwise.
We supplemented the NIKA2 data with 
a dust temperature map at 36.3$^{\prime\prime}$ resolution obtained from HGBS data through spectral energy distribution (SED) fitting 
to four {\it Herschel} bands between 160 and 500\,$\mu$m (\cite{palmeirim13,marsh16}, see http://gouldbelt-herschel.cea.fr/archives).

\section{Dust Emissivity Index Map}

\subsection{Deriving the Dust Emissivity Index $\beta$}
\label{sect:derivebeta}
\begin{figure*}
\begin{tabular}{c}
\includegraphics[width=9.2cm]{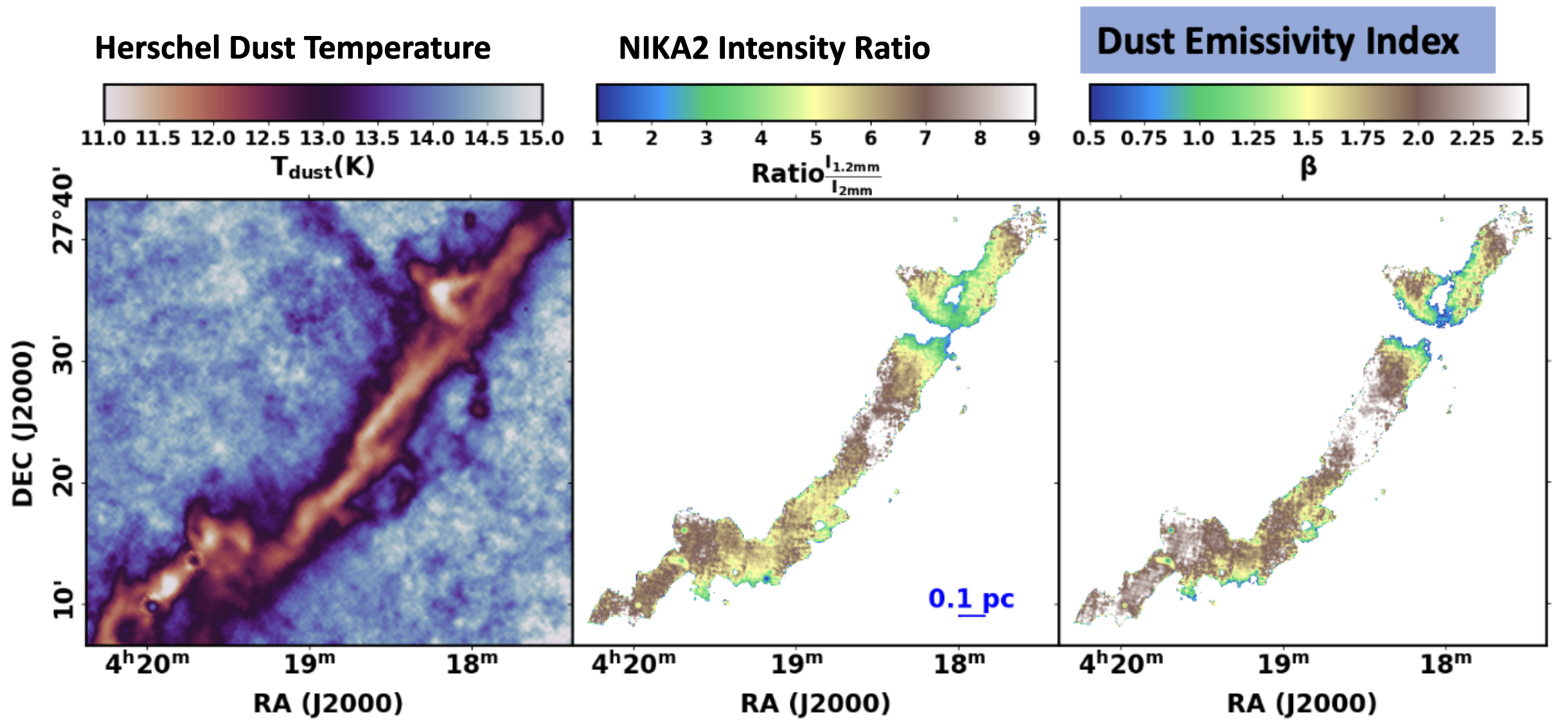}
\end{tabular}
      \caption[]{
        {\bf (a)}: Dust temperature ($T_{\rm d}$)  map derived from {\it Herschel} GBS data \cite{palmeirim13,marsh16}.
        {\bf (b)}: Intensity ratio ($R_{1,2}=\frac{I_{1\rm mm}}{I_{2\rm mm}}$) map derived from the two NIKA2 bands. 
        {\bf (c)}: Dust emissivity index ($\beta$) map derived from the ratio and the temperature map as explained in Sect.\,\ref{sect:derivebeta}.}
\label{fig:betamap}
\end{figure*}

First, we derived a ratio map $R_{1.2\_2} = \frac{I_{\nu_{1.2}} } {I_{\nu_{2}}}$ by dividing the the 1.2\,mm map by the 2.0\,mm map 
on a pixel-by-pixel basis, after smoothing the former to the resolution of the latter and concentrating on those pixels with
high signal-to-noise ratio (SNR > 5) in both maps. 
Then, we smoothed the ratio map to the 36.3$^{\prime\prime}$ resolution of the 
dust temperature map 
and derived the $\beta$ index from the following equation assuming optically thin dust emission:  
$\beta = \log \left( R_{1.2\_2} \frac{B_{\nu_{2}}\left( T_{\rm d}\right)}{B_{\nu_{1.2}}\left( T_{\rm d}\right)} \right) / \log \left( \frac{\nu_{\rm 1.2}}{\nu_{\rm 2}} \right).$
Here, $B_{\nu_{1.2}}\left( T_{\rm d}\right)$ and ${B_{\nu_{2}}\left( T_{\rm d}\right)}$ are the Planck blackbody functions for temperature $T_{\rm d}$  
at 1.2~mm and 2~mm, respectively. $T_{\rm d}$ is given by the {\it Herschel} dust temperature map.
The maps are shown in Fig.~\ref{fig:betamap}.

\subsection{Overall Dust Emissivity Index from Filament to T-Tauri Star}
\label{sec:overal_beta}
\begin{figure}[t]
\vspace{-7mm}
\begin{center}
\sidecaption
\includegraphics[scale=0.4]{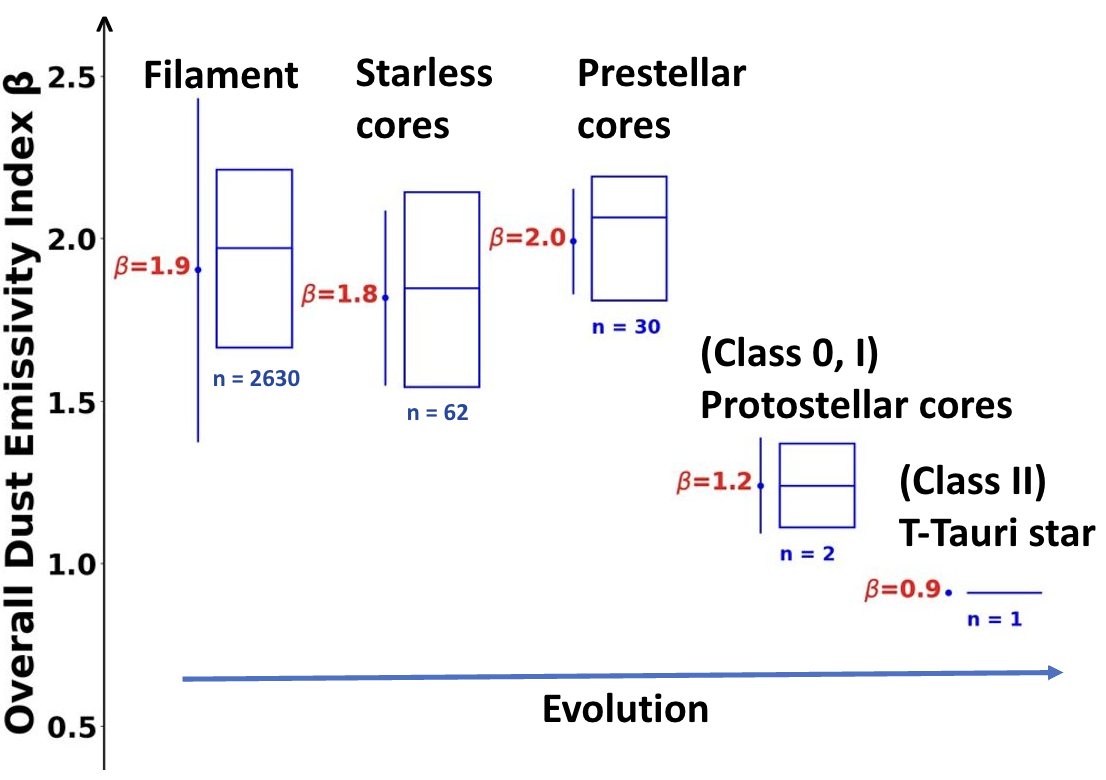}
\label{fig:betastat}
\caption{Overall evolution of $\beta$ from the parent filament to starless cores, prestellar cores, protostellar cores, and T-Tauri disk in the Taurus B211/B213 region. 
The mean $\beta$ value is 
$\sim1.9\pm0.6$ in the filament, $\sim1.8\pm0.3$ in (unbound) starless cores,  $\sim2.0\pm0.2$ in prestellar cores, 
and decreases to $\sim1.2\pm0.2$ in protostellar cores and $\sim0.9\pm0.3$ in T Tauri protoplanetary disks.
The vertical bars mark $1\sigma$ standard deviations around the mean values. 
Also plotted are  
the median and Inter Quartile Range (25\% and 75\%) values for each sub-sample in the partial boxplots. 
The number below each boxplot gives the sub-sample size.}
\end{center}
\vspace{-7mm}
\end{figure}

Using our $\beta$ map and the positions of the compact sources in the dense core catalog derived from 
{\it Herschel} GBS data in Taurus \cite{marsh16, andre10}, we estimated the corresponding $\beta$ values 
averaged over a radius of 0.01~pc from each source center. 
These compact sources are classified as objects at different evolutionary stages: (unbound) starless cores, prestellar cores, protostellar cores, T-Tauri disk.  
By definition, starless cores do not contain any (proto)stellar object; prestellar cores are the subset of starless cores that are gravitationally bound and 
are thus representative of the initial conditions for protostellar collapse \citep{konyves15,marsh16}. 
Protostellar cores are gravitationally collapsing and host a Class~0 or Class~I YSO. 
Class~II YSOs are embedded pre-main sequence (or T Tauri) stars which usually host a protoplanetary disk. 
 As can be seen in Fig.~3, 
 the mean overall $\beta$ index at different evolutionary stages ranges from $\sim$0.9 to $\sim$2.   
Our results for a sample of 105 objects in the B211/B213 filament region indicate that the mean dust emissivity index $\beta$ decreases 
continuously from the parent filament and prestellar cores ($\beta \sim 2\pm0.5$) 
to protostellar cores ($\beta \sim 1.2 \pm 0.2$) to T-Tauri protoplanetary disks ($\beta < 1$). This is consistent with the similar, pioneering study with NIKA-1 by \citep{bracco17} for a much smaller sample of only 5 objects.   
While the numbers of protostellar cores and T-Tauri disks in our sample are very small, 
the low $\beta$ values we find are consistent with the results of both the CALYPSO IRAM PdB survey on 12 protostellar cores at $\sim$500 AU scale 
\citep{galametz19} and an ALMA survey of 36 T-Tauri disks in Lupus at $\sim$50 AU resolution \citep{Tazzari21}.
Overall, our findings suggest that there is a continuous decrease of $\beta$ as star formation proceeds. 
Since a marked decrease in $\beta$ only starts between the prestellar core and the protostellar core stage in our sample, 
we suggest that dust grains begin to grow significantly in size only after the onset of gravitational contraction/collapse from prestellar cores to protostars, 
eventually reaching big sizes in T Tauri protoplanetary disks. 
This interpretation is consistent with model calculations of dust grain growth in star-forming regions due to coagulation \citep{Ossenkopf94, Stognienko95, Henning96}. 

\subsection{Dust Emissivity Index Profile Across the B211/B213 Filament}

\begin{figure*}[tbh]
\centering
\begin{tabular}{c}
\includegraphics[width=11.5cm]{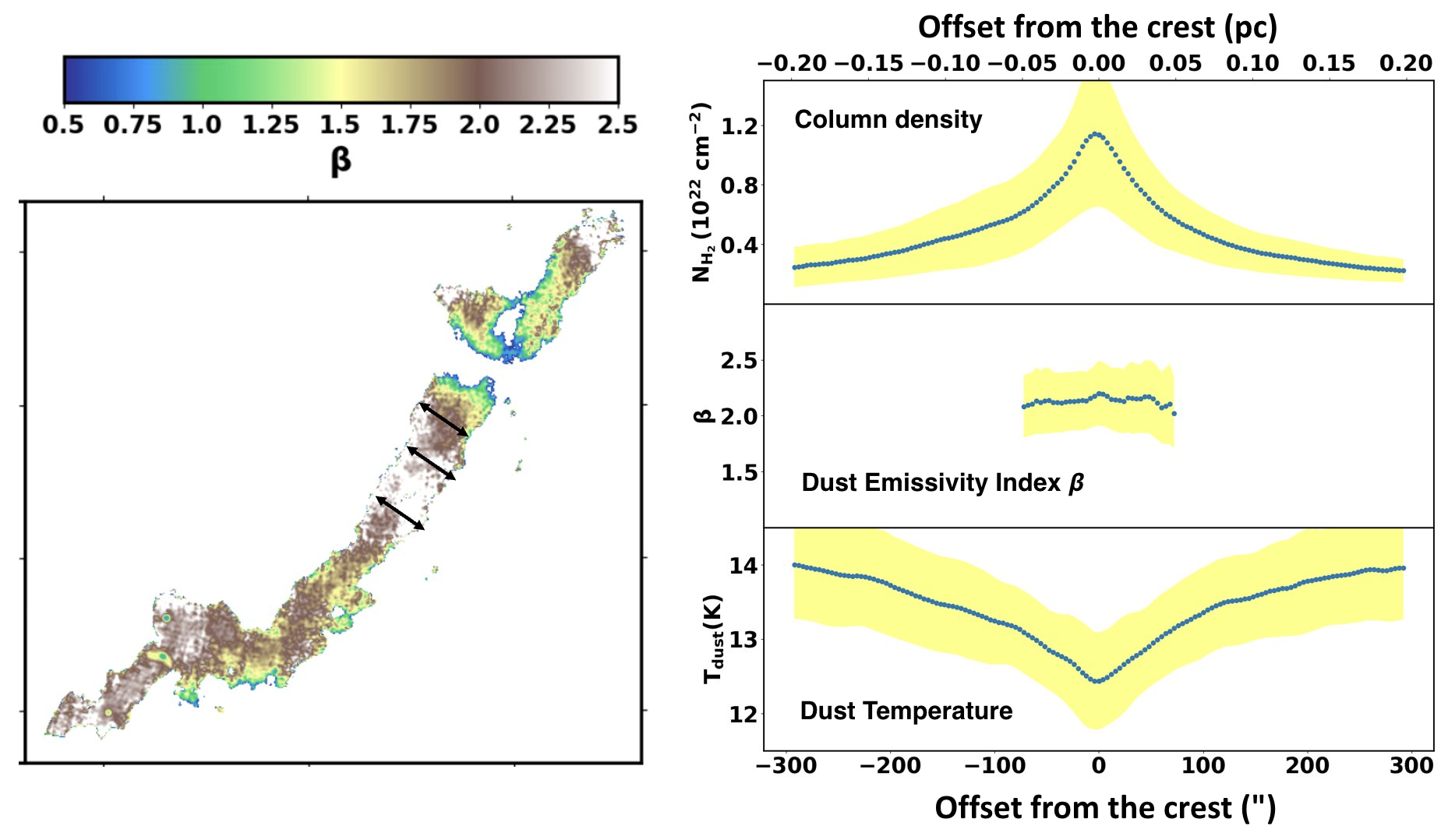}
\end{tabular}
      \caption[]{
      {\bf (Left)}: Map of the dust emissivity index $\beta$ in Taurus B211/B213, with several examples of radial cuts across the filament.
      {\bf (Right)}: Mean radial column density profile (top), mean radial profile of the dust emissivity index $\beta$ (middle), 
      and mean dust temperature profile (bottom) across the B211/B213 filament. The yellow-shaded areas show $1\sigma$ deviations 
      around the mean values.}
\label{fig:betaprofile}
\vspace*{-0.7cm}
\end{figure*}

Our high-sensitivity intensity NIKA2 maps allow us to construct a $\beta$ map for the entire B211/B213 filament (Fig.~\ref{fig:betamap}c)  
and to derive an average radial profile of the $\beta$ index across the filament. 
Using the filament crest derived from the {\it Herschel} GBS column density map \cite{palmeirim13,marsh16}, 
we derived $\beta$ profiles along 70 cuts perpendicular to the crest out to the filament boundary at  steps of 36$^{\prime\prime}$. 
The mean $\beta$ profile is shown in Fig.\,\ref{fig:betaprofile}, along with the column density and dust temperature profiles. 
While $\beta$ exhibits local variations from 0.5 to 2.5 as seen in the $\beta$ map (Fig.~\ref{fig:betamap}c), the mean transverse profile of $\beta$ 
averaged along the filament crest is flat with a typical value $\beta\sim2\pm0.3$ (Fig.~\ref{fig:betaprofile}-Right).  
This differs from the results of \citep{howard19}, who derived an average $\beta$ value of $<$\,1.5 in the filament 
and found significant variations of $\beta$ across the filament. 
The $\beta$ estimates of \citep{howard19} were obtained by adding a SCUBA2 850\,$\mu$m data point to 
the {\it Herschel} measurements between 70 and 500\,$\mu$m and used a sophisticated SED fitting scheme, 
but suffered from the lack of data points in the Rayleigh-Jeans domain.
We argue that our NIKA2 measurements at 1.2 and 2.0\,mm provide better, more direct constraints on $\beta$ and are more reliable 
as they are made at longer wavelengths, probing the Rayleigh-Jeans portion of the SED where temperature effects are minimized. 
We conclude that the column density map of the B211/B213 region and density profile of the filament derived 
by \cite{palmeirim13} from HGBS data assuming a constant $\beta =2$ value (consistent with our present $\beta\sim2\pm0.3$ estimate) 
are most likely reliable and more robust than the maps/profiles derived by \citep{howard19} 
with a varying, uncertain $\beta$.

\section*{Acknowledgements} \label{ack}
\vspace{-1mm}
We would like to thank the IRAM staff for their support during the observation campaigns. The NIKA2 dilution cryostat has been designed and built at the Institut N\'eel. In particular, we acknowledge the crucial contribution of the Cryogenics Group, and in particular Gregory Garde, Henri Rodenas, Jean-Paul Leggeri, Philippe Camus. This work has been partially funded by the Foundation Nanoscience Grenoble and the LabEx FOCUS ANR-11-LABX-0013. This work is supported by the French National Research Agency under the contracts "MKIDS", "NIKA" and ANR-15-CE31-0017 and in the framework of the "Investissements d’avenir” program (ANR-15-IDEX-02). This work has benefited from the support of the European Research Council Advanced Grant ORISTARS under the European Union's Seventh Framework Programme (Grant Agreement no. 291294). A. R. acknowledges financial support from the Italian Ministry of University and Research - Project Proposal CIR01$\_00010$. S. K. acknowledges support provided by the Hellenic Foundation for Research and Innovation (HFRI) under the 3rd Call for HFRI PhD Fellowships (Fellowship Number: 5357). 
Q. Nguyen-Luong and A. Duong-Tuan were partly supported by a grant from the Simons Foundation (916424, N.H.) in addition to the enthusiastic support of IFIRSE/ICISE staff.

\bibliography{tau,tau2}

%
%
%







\end{document}